\documentclass[aps,twocolumn,floatfix,altaffilletter,superscriptaddress,preprintnumbers,tightenlines,showpacs,showkeys,notitlepage,nofootinbib]{revtex4-2}

\usepackage[colorlinks=true,citecolor=blue,linkcolor=blue]{hyperref}
\usepackage[normalem]{ulem}
\usepackage{amsmath,amssymb}
\usepackage{mathtools}
\usepackage{verbatim}
\usepackage{titlesec}               
\usepackage{epsfig}
\usepackage{graphicx}               
\usepackage{url}
\usepackage{color}
\usepackage{multirow}
\usepackage{placeins}
\usepackage[dvipsnames]{xcolor}
\usepackage{epstopdf}
\usepackage{siunitx}
\usepackage{fontawesome}
\usepackage{enumitem}
\usepackage[capitalize]{cleveref}
\usepackage{lipsum}
\usepackage{gensymb}
\usepackage{aas_macros}             
\usepackage{booktabs}               
\usepackage{xspace}                 
\usepackage{pifont}
\usepackage{marvosym }
\allowdisplaybreaks

\makeatletter
\renewcommand{\thesection}{\arabic{section}}

\renewcommand{\p@subsection}{}
\makeatother

\titleformat*{\section}{\centering\bfseries\uppercase}
\titlelabel{\thetitle\quad}
\titleformat*{\paragraph}{\bfseries}
\titlespacing*{\paragraph}{0pt}{3.25ex plus 1ex minus .2ex}{1em}

\makeatletter
\def\l@subsubsection#1#2{}
\makeatother

\begin{document}
\title{
Improving Neutrino Energy Reconstruction with Machine Learning
}

\author{Joachim Kopp}
\email{jkopp@cern.ch}
\affiliation{Theoretical Physics Department, CERN,
             1 Esplanade des Particules, 1211 Geneva 23, Switzerland}
\affiliation{PRISMA Cluster of Excellence \& Mainz Institute for
             Theoretical Physics, \\
             Johannes Gutenberg University, Staudingerweg 7, 55099
             Mainz, Germany}

\author{Pedro Machado}
\email{pmachado@fnal.gov}
\affiliation{Particle Theory Department, Fermilab, P.O. Box 500, Batavia, IL 60510, USA}

\author{Margot MacMahon}
\email{margot.macmahon.21@ucl.ac.uk}
\affiliation{University College London, London, WC1E 6BT, United Kingdom}

\author{Ivan Martinez-Soler}
\email{ivan.j.martinez-soler@durham.ac.uk}
\affiliation{Institute for Particle Physics Phenomenology, Department of Physics, Durham University, South Road, Durham, U.K.}

\date{\today}
\preprint{CERN-TH-2024-066, MITP-24-052, FERMILAB-PUB-24-0276-T, IPPP/24/26}

\begin{abstract}
\noindent
Faithful energy reconstruction is foundational for precision neutrino experiments like DUNE, but is hindered by uncertainties in our understanding of neutrino--nucleus interactions.
Here, we demonstrate that dense neural networks are very effective in overcoming these uncertainties by estimating inaccessible kinematic variables based on the observable part of the final state.
We find improvements in the energy resolution by up to a factor of two compared to conventional reconstruction algorithms, which translates into an improved physics performance equivalent to a 10--30\% increase in the exposure.
\end{abstract}

\maketitle

\textbf{Introduction.---}Current and upcoming neutrino experiments are embarking on an extensive program of precision measurements, which may have far-reaching ramifications for our understanding of the Universe.
One of the major goals of accelerator-based long-baseline oscillation experiments like NO$\nu$A~\cite{NOvA:2021nfi}, T2K~\cite{T2K:2023smv}, DUNE~\cite{DUNE:2020ypp}, and Hyper-Kamiokande~\cite{Hyper-Kamiokande:2018ofw} is the discovery of leptonic $CP$ violation, which could be crucial for explaining the puzzling matter--antimatter asymmetry of the Universe.
Together with the upgraded IceCube telescope~\cite{Ishihara:2019aao}, ORCA~\cite{KM3NeT:2021ozk}, and JUNO~\cite{JUNO:2015sjr}, these experiments will also be sensitive to the neutrino mass ordering (an important input to neutrinoless double beta decay experiments aiming to probe the Dirac or Majorana nature of neutrinos) \cite{Kelly:2020fkv}, the octant of the mixing angle $\theta_{23}$, and a host of other parameters related to neutrino mixing and to the physics of neutrino--nucleus interactions.
Common to all these measurements is their reliance on accurate neutrino energy reconstruction, which is the topic of this work.

Our focus will be on liquid argon time projection chambers (LArTPCs) such as DUNE. 
These detectors exhibit exceptional event reconstruction capabilities, opening exciting new avenues for measurements both within the Standard Model~\cite{Kelly:2019itm, Ternes:2019sak, Kelly:2021jfs} and beyond~\cite{deGouvea:2018cfv, DeRomeri:2019kic, Harnik:2019zee, Berryman:2019dme, Breitbach:2021gvv, Bertuzzo:2022fcm}. 
A decade ago, ArgoNeuT demonstrated that LArTPCs are able to reconstruct protons with an energy threshold of only \SI{21}{MeV}~\cite{ArgoNeuT:2014ihi}.
More recently, MicroBooNE has presented an impressive set of analyses, including for instance a sensitive measurement of radiative $\Delta(1232)$ decays, which are a crucial background to oscillation analyses~\cite{MicroBooNE:2021zai}.

\begin{figure}
    \centering
    \includegraphics[width=0.7\columnwidth]{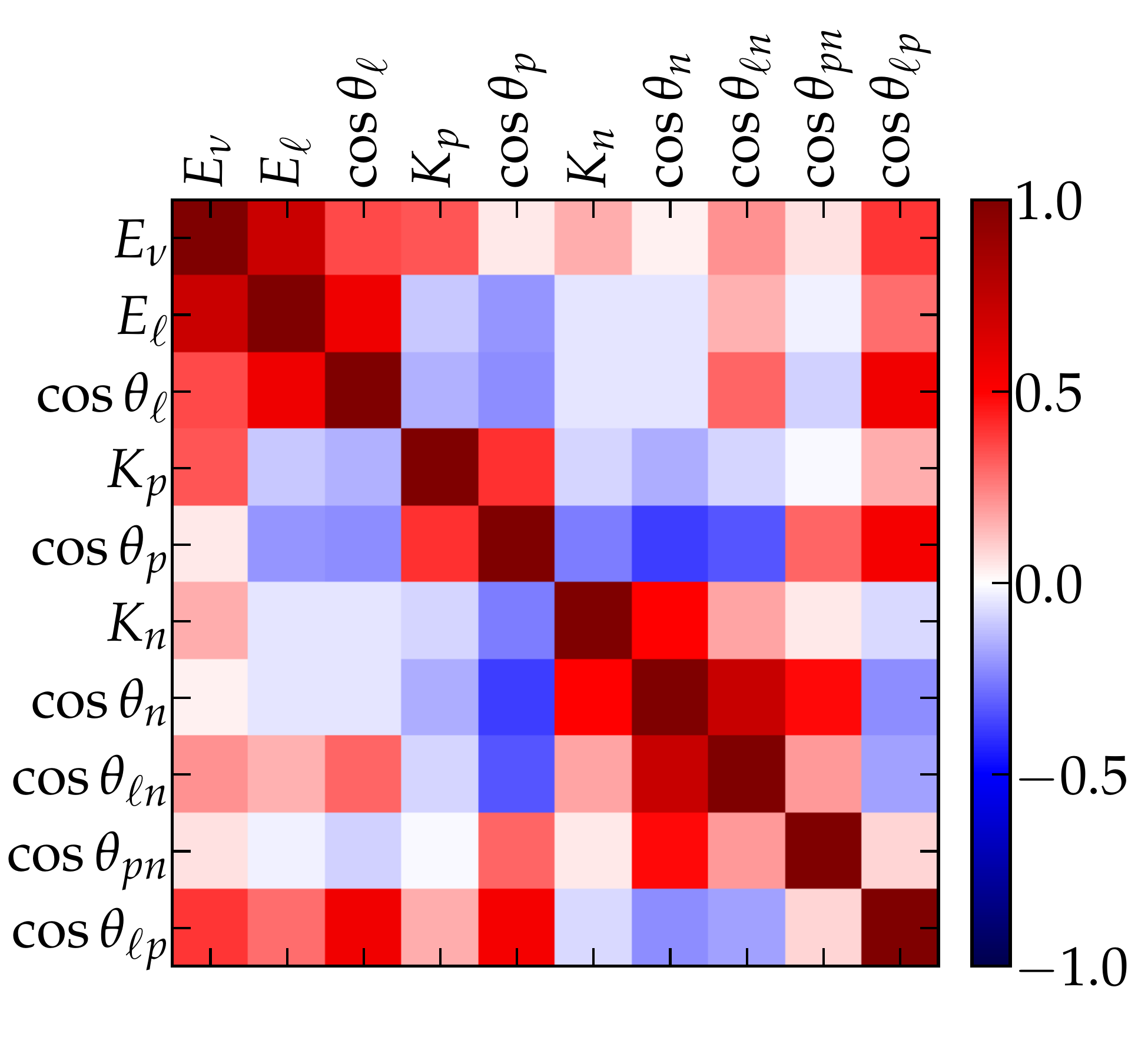}
    \caption{Example of a Pearson correlation matrix between several observables in charged-current neutrino--argon interactions with at least one proton and one neutron in the final state, for the DUNE neutrino-mode flux and before including detector responses.
    We define the proton ($p$) and neutron ($n$) systems by adding up their kinetic energies and three-momenta.
    We include the total kinetic energy of the proton and systems ($K_p$, $K_n$), the energies of the neutrino and outgoing lepton ($E_\nu$, $E_\ell$), the directions $\cos\theta_{\ell,p,n}$ relative to the beam axis, and the opening angles $\cos\theta_{\ell p,\ell n, pn}$.
    Note that neutrons are very challenging to reconstruct, so information on the neutron system is typically not available in realistic event records.}
    \label{fig:correlations}
\end{figure}

Our goal is to demonstrate how the abundance of information contained in a LArTPC neutrino event can be leveraged to significantly improve the reconstruction of the incoming neutrino on an event-by-event basis.
In fact, intranuclear effects in the neutrino interaction lead to important nontrivial correlations among the incoming neutrino energy, $E_\nu$, and different kinematic variables characterizing the final state, see \cref{fig:correlations}.
Understanding these correlations is crucial because not all kinematic variables (especially those related to final-state neutrons) can be reconstructed.
But doing so analytically is extremely challenging.
The problem thus calls for machine learning techniques, which are well suited for dealing with correlations in high-dimensional parameter spaces.

We will estimate how such techniques can improve the determination of the neutrino energy for both beam neutrinos and atmospheric neutrinos. 
For the latter, we will also study the reconstruction of the incoming neutrino direction, which  translates into the distance the neutrino has travelled -- a key ingredient in any oscillation analysis.
It is particularly important for sub-GeV neutrinos which can give DUNE sensitivity to $CP$ violation before the beam turns on~\cite{Kelly:2019itm}, and which allow for neutrino tomography of the Earth's interior~\cite{Kelly:2021jfs, Denton:2021rgt}.
Besides DUNE, our results can also benefit the LArTPCs comprising Fermilab's Short Baseline Neutrino Program~\cite{MicroBooNE:2015bmn}.

\textbf{Neutrino energy and angle reconstruction.---}To estimate how much we can improve the neutrino energy reconstruction in LArTPCs, we proceed as follows.
We generate 800,000 $\nu_\mu$--argon charged current events in NuWro~$21.09$~\cite{Golan:2012rfa}, with the neutrino energies distributed according to the DUNE $\nu_\mu$ flux in the forward horn current (``neutrino mode'') configuration~\cite{DUNE:2020ypp}.
We model detector effects by imposing a kinetic energy threshold for charged particles, and by smearing their angles and momenta -- see~\cref{tab:detection} for details. 
We do not assume any charge identification capabilities.
For neutrons, we consider three different cases: (i) no reconstruction at all, labeled ``$0_n$''; (ii) reconstruction of the neutron energy with a fractional resolution of 
$40\%/\sqrt{K_n/\si{GeV}}$, ``$E_n$''; and (iii) reconstruction of the neutron energy and direction, with a resolution of $10^\circ$ for the latter, ``$E_n+\theta_n$.''
Neutrons propagating in liquid argon leave multiple small deposits of energy, or blips, in the detector~\cite{Friedland:2018vry}, and while such blips have been detected \cite{ArgoNeuT:2018tvi}, actual neutron reconstruction has not been firmly demonstrated.
The three scenarios we consider may serve as further motivation to focus effort on this challenging task.
To simulate atmospheric neutrinos, we randomize the incoming neutrino direction.

\begin{table}[t]
    \begin{center}
        \begin{tabular}{cccc}
        \midrule
        Particle             & Threshold     & resolution $\alpha_p$  & $\sigma(\theta)$ \\
        \midrule
        $\mu$, $e$, $\gamma$ & \SI{30}{MeV}  &  5\%  & $ 2^\circ$ \\ 
        $\pi$, $K$, proton   & \SI{30}{MeV}  & 10\%  & $10^\circ$ \\ 
        \midrule
        neutron ``$0_n$''       & invisible     & --   & -- \\ 
        neutron ``$E_n$''       & \SI{100}{MeV} & 40\% & -- \\ 
        neutron ``$E_n$+$\theta_n$'' & \SI{100}{MeV} & 40\% & $10^\circ$ \\
        \midrule
        \end{tabular}
    \end{center}
    \caption{Kinetic energy threshold, momentum resolution ($\sigma(p) = \alpha_p\sqrt{p/\si{GeV}}$), and angular resolution for different final-state particles. For neutron reconstruction, we list the three different scenarios.}
    \label{tab:detection}
\end{table}

Before passing events to a neural network, we combine all particles of the same species above a certain threshold into a single entity, summing over their energies and three-momenta.
This makes our results more robust with respect to modelling uncertainties in neutrino event generators.
For example, GENIE~\cite{Andreopoulos:2009rq} and NuWro~\cite{Golan:2012rfa} exhibit large discrepancies in the number of predicted low-energy protons in the final state.

\begin{figure}
    \centering
    \includegraphics[width=\columnwidth]{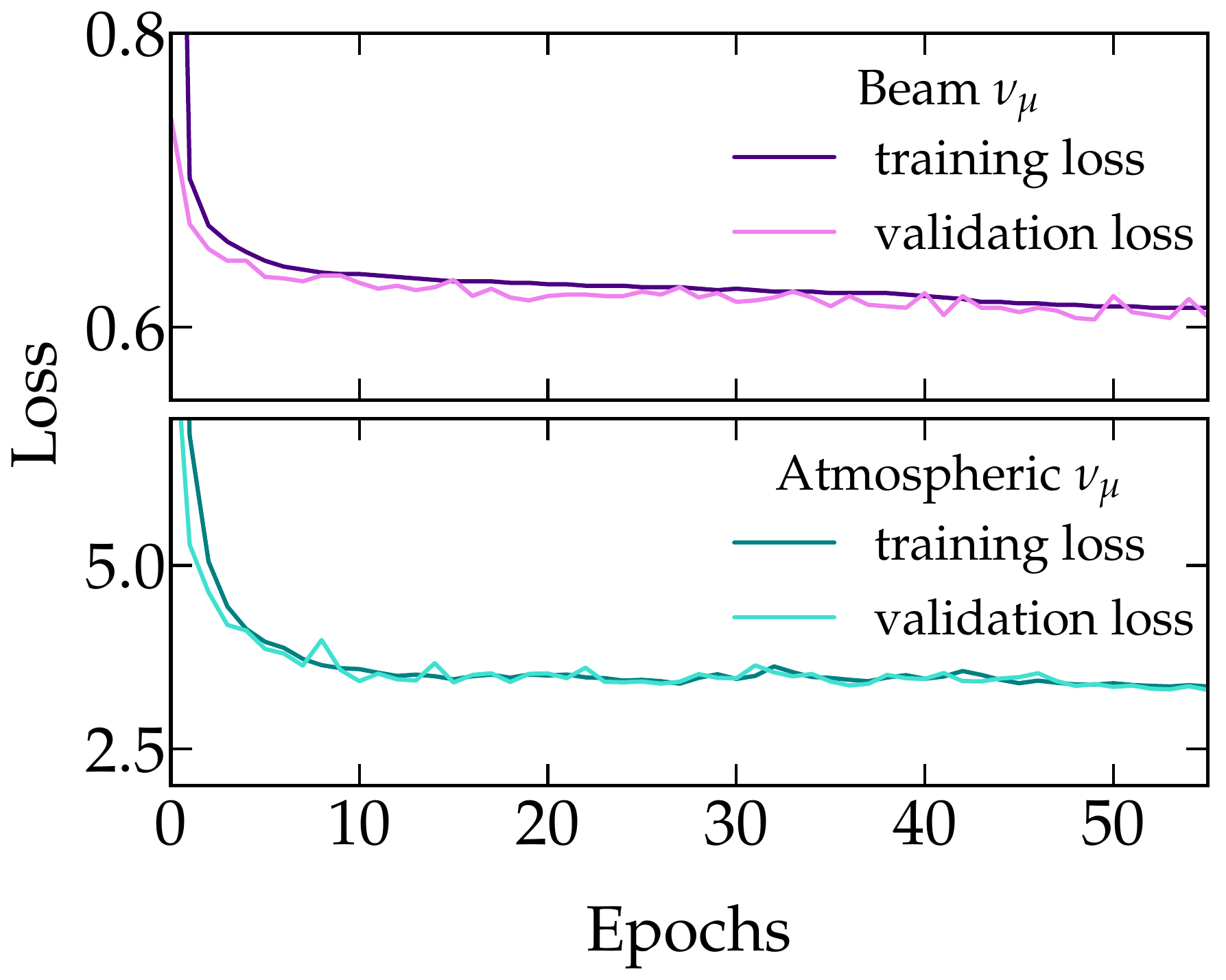}
    \caption{DNN loss functions when trained on beam neutrino events (upper panel) and atmospheric neutrino events (lower panel) for training (darker) and validation (brighter) data. These results are for the ``0n'' (no neutron reconstuction) scenario, but results are similar for the ``En'' and ``$E_n + \theta_n$'' cases.}
    \label{fig:losses}
\end{figure}

We use dense neural network (DNN) classifiers, with the true neutrino energy as the label for beam neutrinos, and the energy, zenith and azimuth angles for atmospheric neutrinos.
The DNN consists of an input layer and two dense hidden layers plus one dense output layer, with 64--16--32--1 nodes, respectively.
In \cref{fig:losses}, we present the loss function for training and validation data. 
For beam neutrinos we define the loss function as the fractional mean squared error $L_\text{energy} = 10\left[1 - E_{\nu,\theta}(x)/E_\nu(x)\right]^2$, where $E_\nu(x)$ ($E_{\nu,\theta}(x)$) denotes the true (reconstructed) neutrino energy for input variables $x$ and DNN parameters $\theta$; while for atmospheric neutrinos we add $L_\text{angle} = 30\arccos^{2}\left[\hat{v}_\theta(x) \cdot \hat{v}(x) \right]$, where $\hat{v}$ ($\hat{v}_\theta(x)$) is a unit vector in the true (reconstructed) direction of the incoming neutrino momentum.

\begin{figure}
    \centering
    \includegraphics[width=\columnwidth]{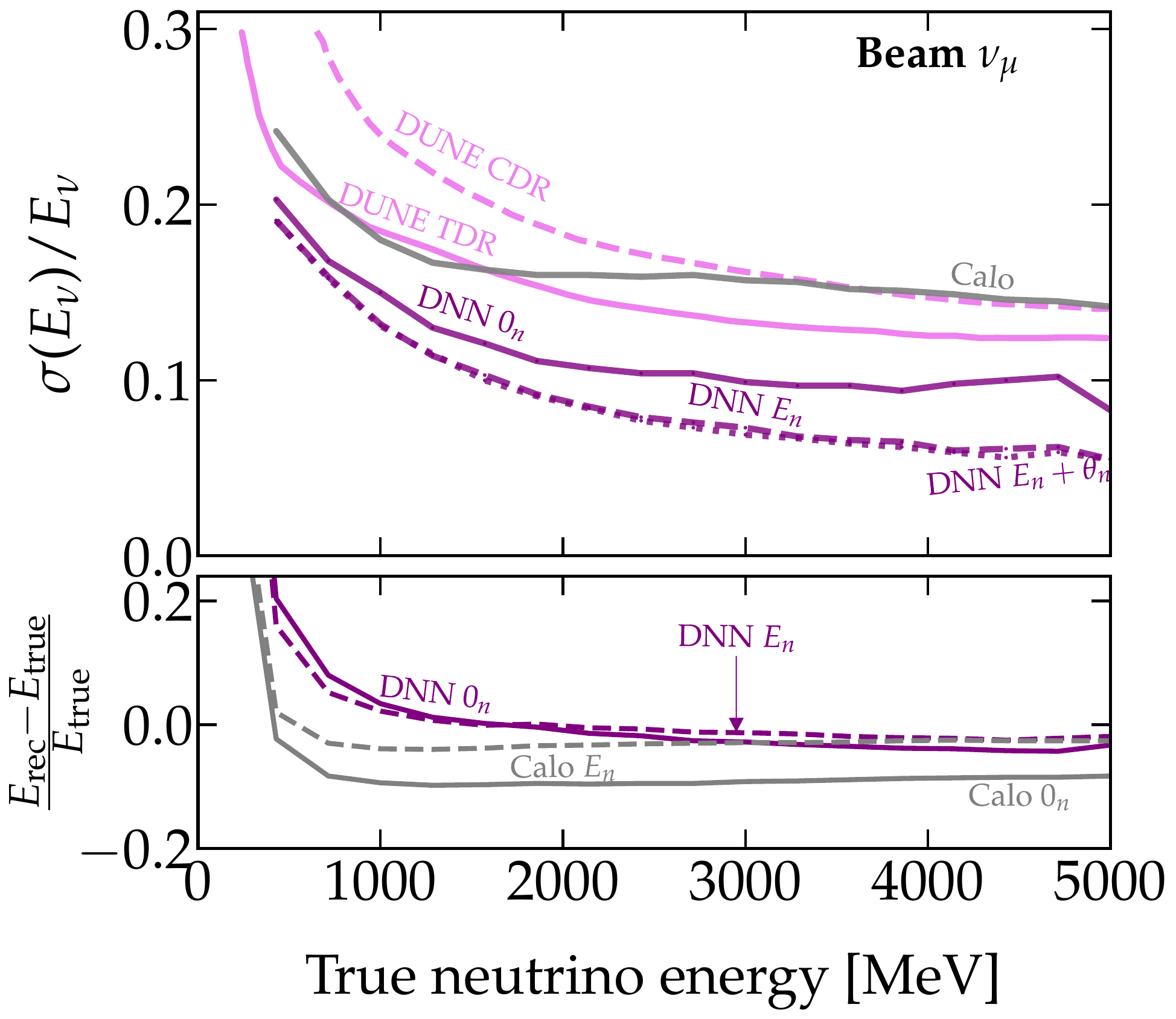}
    \caption{\emph{Top:} fractional neutrino energy resolution $\sigma(E_\nu)/E_\nu$ as a function of neutrino energy from DNN-based analyses with no information on final-state neutrons (solid purple), with limited information on the neutron energy (dashed purple), and with information on the neutron energy and direction (dotted purple).
        For comparison, we also show the energy resolutions anticipated in the DUNE CDR and TDR simulations (magenta), and the resolution of a simple calorimetric method assuming invisible neutrons (gray).
        \emph{Bottom:} reconstruction bias for the DNN compared to the calorimetric method.
        \emph{DNNs significantly outperform conventional approaches to energy reconstruction.
        When information on the energy of final-state neutrons is available, the improvement is more than a factor of two at high energies.}}
    \label{fig:beam}
\end{figure}

The performance of the DNN for DUNE beam data can be seen in \cref{fig:beam}, where we present the fractional energy resolution as a function of neutrino energy.
We show results for the three neutron reconstruction scenarios (purple lines) and compare to the anticipated energy resolutions quoted in DUNE's Conceptual Design Report (CDR)~\cite{DUNE:2016ymp} and Technical Design Report (TDR)~\cite{DUNE:2021cuw}. 
We also compare to a purely calorimetric method (gray), in which the neutrino energy is obtained as 
\begin{align}
    E^\text{cal}_\nu = E_\ell + \sum_i^{\rm mesons} E_i + \sum_i^{\rm baryons} K_i,
\end{align}
where $E_i$ and $K_i$ denote the total and kinetic energy of a particle $i$, respectively.
We observe that the performance of the calorimetric method using our simplified detector simulation falls between the CDR and TDR methods.
Our main result is the observation that the DNN improves the neutrino energy resolution, $\sigma(E_\nu)$, considerably at all energies.
This is true even when neutrons are completely invisible because the DNN is able to partially infer the kinematics of final-state neutrons based on correlations with other kinematic variables alone.
If the neutron energy can be reconstructed (``$E_n$''), even with only the 40\% resolution assumed here, $\sigma(E_\nu)$ improves further by a significant amount.
Adding information on the neutron direction (``$E_n + \theta_n$''), however, does not lead to additional improvements.
Results for the reconstruction of antineutrinos are similar.
Since our simulation treats the detector response to muons and electrons similarly, our conclusions for electron neutrino events are essentially the same as for the muon neutrino events shown here.

\begin{figure}
    \centering
    \includegraphics[width=\columnwidth]{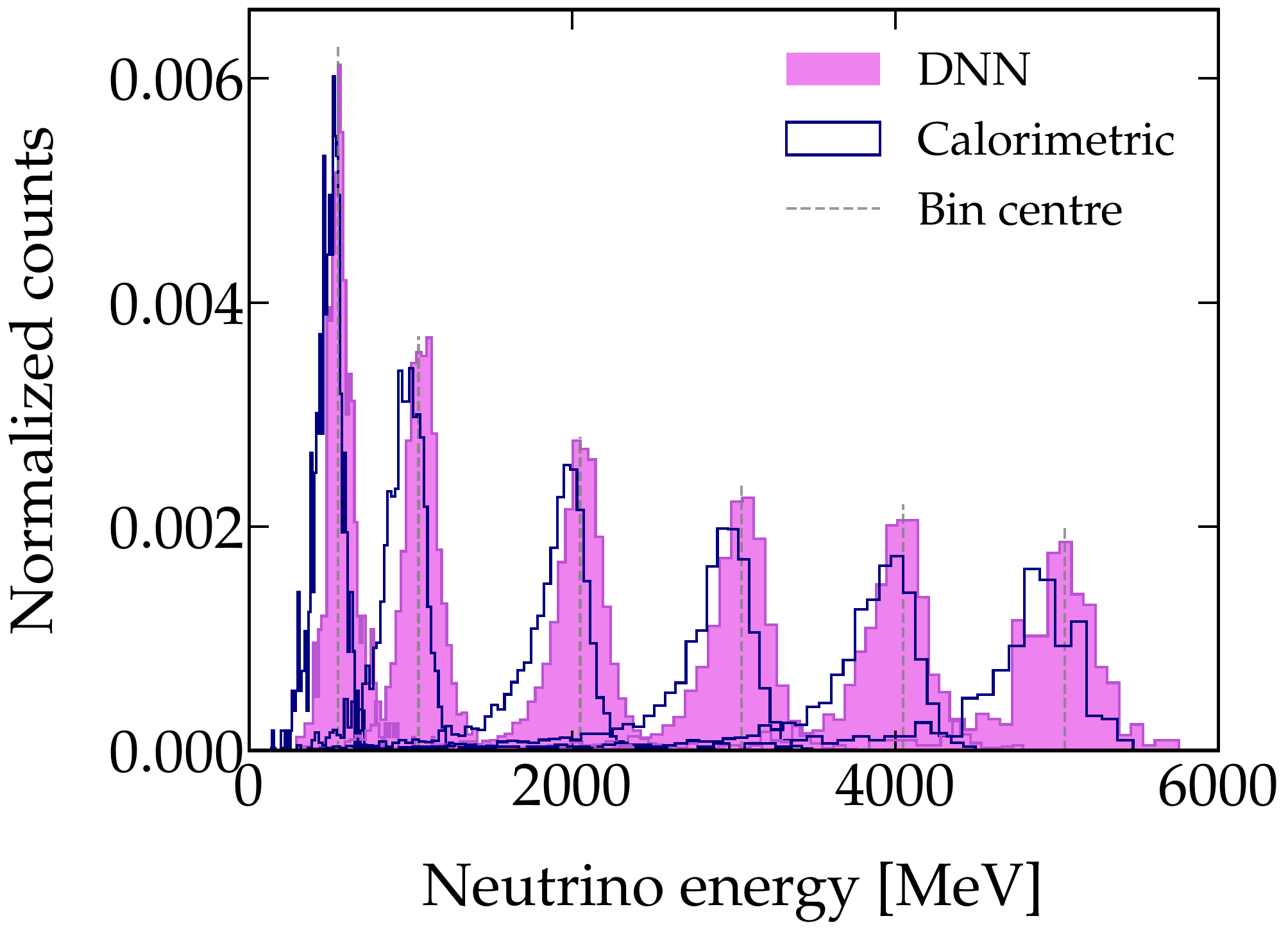}
    \caption{Normalized event distribution using DNN energy reconstruction (filled histogram) and the calorimetric method (unfilled histogram). Neutrinos with true energies across six ranges, $[500-600, 1000-1100, 2000-2100, 3000-3100, 4000-4100, 5000-5100  \ \mathrm{MeV}]$, were used. The DNN reconstruction produces a narrower event distribution and reduces the bias in the mean reconstructed energy compared to the calorimetric method.}
    \label{fig:enedist}
\end{figure}

 The improvement in energy reconstruction achieved by the DNN over the calorimetric method is illustrated in \cref{fig:enedist}, which shows a normalized event distribution for neutrinos with true energies across six ranges: $[500-600, 1000-1100, 2000-2100, 3000-3100, 4000-4100, 5000-5100 \ \mathrm{MeV}]$. Events were simulated according to the beam flux, yielding event counts of $[490, 1950, 4320, 4390, 1450, 310]$ per energy range. \Cref{fig:enedist} compares event distributions obtained via the calorimetric method (unfilled histogram) and those reconstructed by the DNN (filled histogram). The DNN reconstruction shows a marked improvement in mean reconstructed energy, significantly reducing the bias inherent in the calorimetric method.  Additionally, the neural network produces a narrower distribution, where the long tails that extend to lower energies visible in the calorimetric approach, are reduced. 

\begin{figure}
    \centering
    \includegraphics[width=\columnwidth]{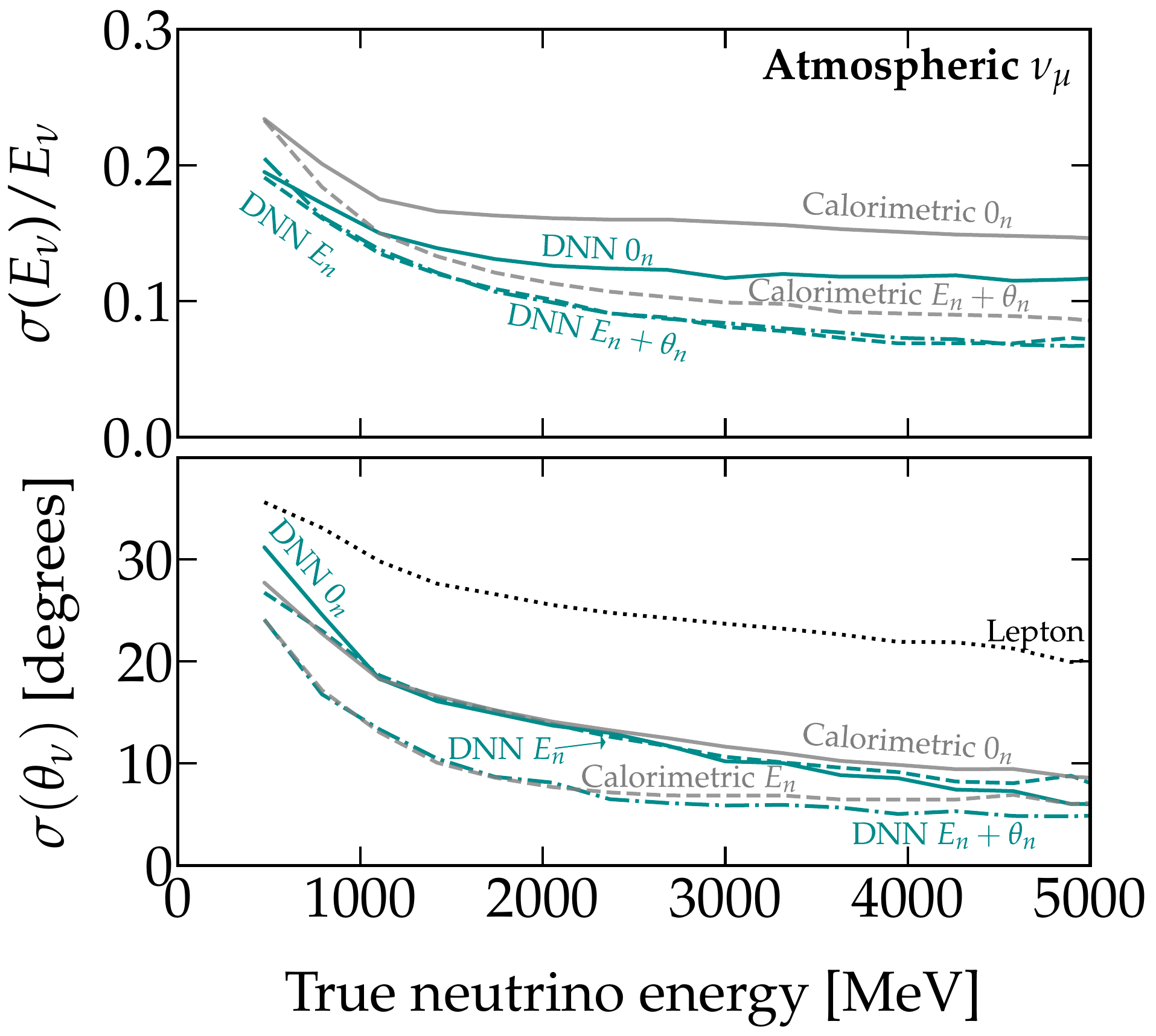}
    \caption{Fractional neutrino energy resolution, $\sigma(E_\nu)/E_\nu$ (top), and angular resolution, $\sigma(\theta_\nu)$ (bottom) from DNN-based analyses of atmospheric neutrino events with no information on final-state neutrons (solid cyan), and with information on neutron energies and directions included (dashed/dot-dashed cyan).
        For comparison, we also show the resolutions achievable with simple calorimetric methods (gray curves).
        The dotted black curve in the bottom panel is based on only the charged lepton kinematics, as in Cherenkov detectors at low energy.}
    \label{fig:atmospheric}
\end{figure}

Going from beam neutrinos to atmospheric neutrinos, where not only the neutrino energy but also the arrival direction need to be reconstructed, we see from \cref{fig:atmospheric} that the DNN again improves $\sigma(E_\nu)$, but not the angular resolution, $\sigma(\theta_\nu)$.
The DNN without information on neutrons achieves a $\sigma(E_\nu)$ closer to the calorimetric method \emph{with} such information included, confirming that the network is able to partially infer the neutron kinematics based on correlations with other kinematic variables.
For $\sigma(\theta_\nu)$ (lower panel of \cref{fig:atmospheric}), the DNN affords only a marginal improvement in the angular resolution when compared to calorimetric methods,\footnote{The calorimetric neutrino direction is given by the vector sum of the three-momenta of all visible outgoing particles.} except at higher energies.
In events with a small hadronic energy fraction, there is simply not enough information to improve neutrino reconstruction any further.
To achieve better results at low energies, an interesting direction of future work could be a DNN-based classifier that separates events into several samples based on the fidelity of the reconstruction \cite{Kelly:2019itm, NOvA:2021nfi}.
In addition, beam experiments such as the SBN detectors and DUNE-PRISM, in which the arrival direction of neutrinos is known, could allow for data-driven development of accurate angular reconstruction methods~\cite{Kelly:2019itm}.

\textbf{Impact on oscillation analyses.---}To emphasize the importance of the improved energy reconstruction afforded by our DNN, we have simulated DUNE's sensitivity to the 3-flavor oscillation parameters using GLoBES \cite{Huber:2004ka, Huber:2007ji, DUNE:2021cuw}.
We compare results obtained with the energy resolution function derived from our DNN without neutron information, approximated as a Gaussian, to results based on the detector response from DUNE's CDR and TDR \cite{DUNE:2020ypp, DUNE:2021cuw}.
The fit and statistical analysis are performed using a code originally developed in Refs.~\cite{Kopp:2013vaa, Dentler:2018sju, github-nuosc}.

\begin{figure}
    \centering
    \includegraphics[width=0.95\columnwidth]{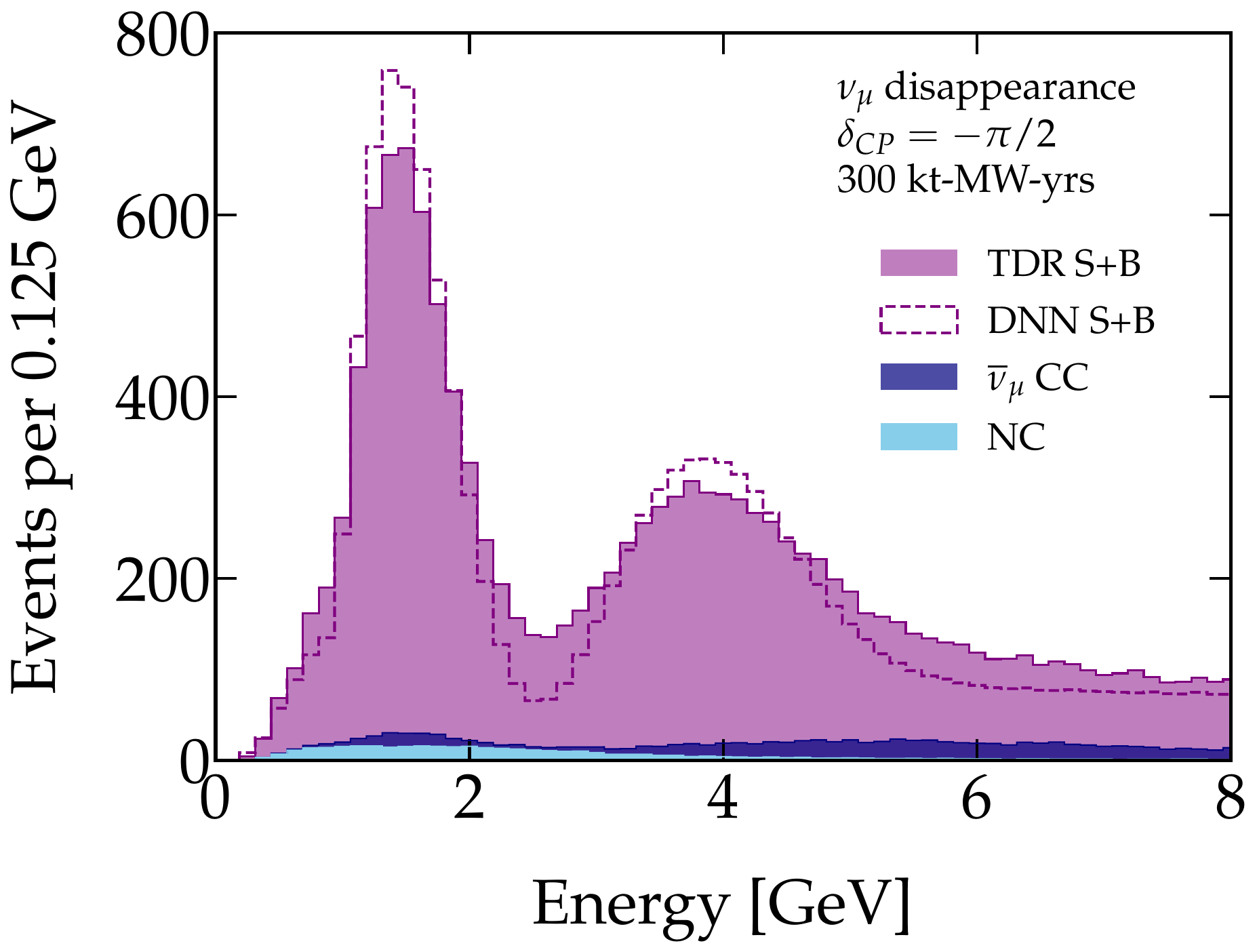}
    \caption{Impact of neural network-improved energy resolution on the energy spectrum of $\nu_\mu$ events in DUNE. Here, $S$ and $B$ denote signal and background.}
    \label{fig:spectra}
\end{figure}

\begin{figure}
    \centering
    \includegraphics[width=0.95\columnwidth]{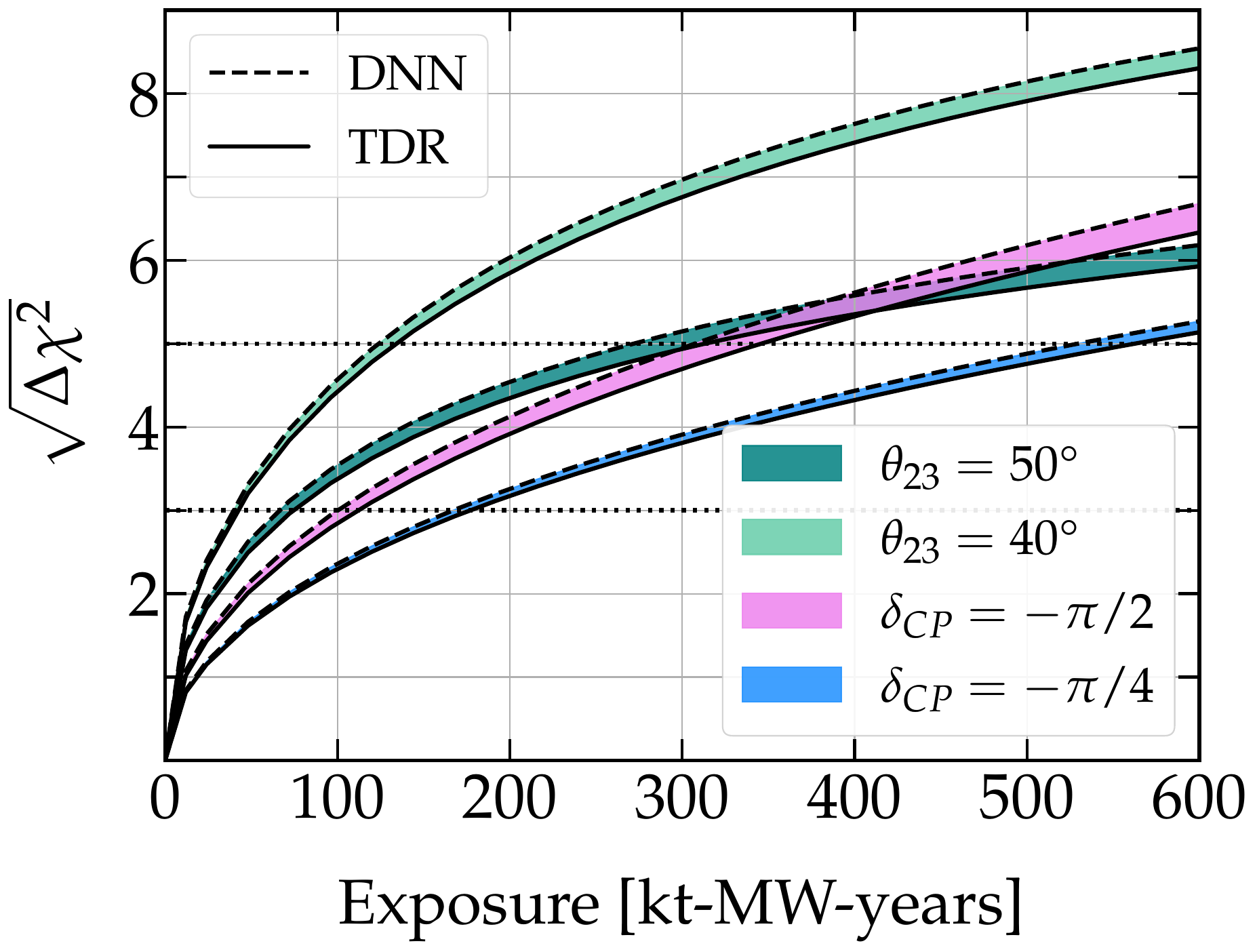}
    \caption{Impact of neural network-improved energy resolution on precision oscillation measurements in DUNE. For the sensitivity to CP violation and to non-maximal $\theta_{23}$, \emph{the improvement in sensitivity due to the DNN is equivalent to a $\sim 10\%$ increase in exposure.}}
    \label{fig:sensitivities}
\end{figure}

\begin{figure}
    \centering
    \includegraphics[width=\columnwidth]{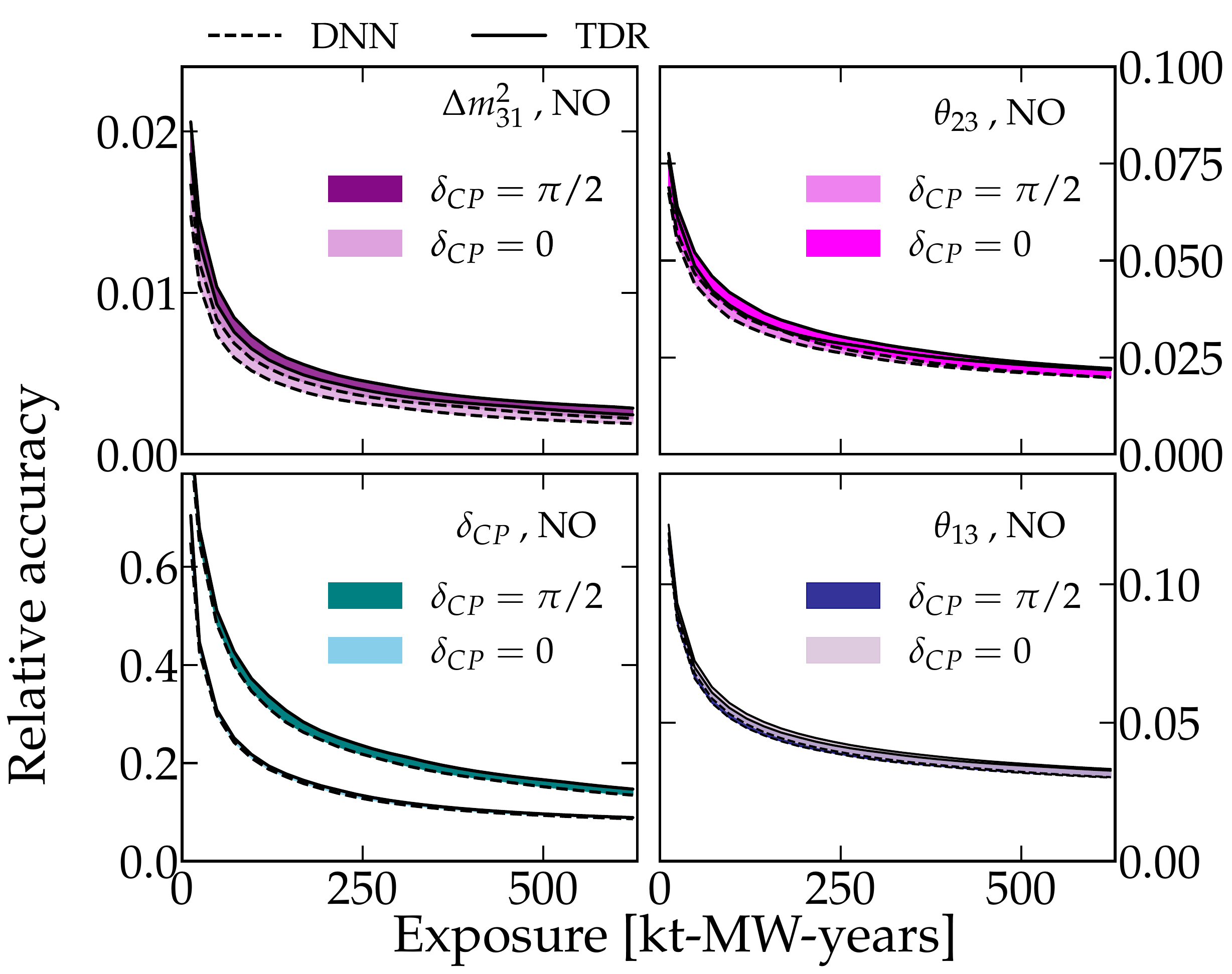}
    \caption{Impact of neural network-improved energy resolution on precision measurements of $\Delta m^2_{31}$, $\theta_{23}$, $\delta_{CP}$, and $\theta_{13}$  in DUNE.}
    \label{fig:relative_precision}
\end{figure}

In \cref{fig:spectra} we show how the improved energy resolution of a DNN-based analysis affects neutrino spectra in DUNE.
Using the $\nu_\mu$ event sample for illustration, the plot shows in particular that oscillations become significantly more pronounced.
Translating this improvement into sensitivities to oscillation parameters leads to the main results of this sensitivity study, shown in \cref{fig:sensitivities,fig:relative_precision}.
We see that precision measurements of the atmospheric oscillation parameters 
-- including searches for deviations of $\theta_{23}$ from maximal mixing, with comparison to $\theta_{23} = 40^{\circ} \; \text{and}\; 50^{\circ}$, 
and the hunt for leptonic $CP$ violation, particularly for cases where $\delta_{CP} = -\pi/2\; \text{and} \; -\pi/4$ - all stand to benefit significantly from DNN-based reconstruction.
The improvement on the $CP$ phase sensitivity for $\delta_{CP} = -\pi/2$ is roughly equivalent to a $\sim 10\%$ increase in exposure -- that is, almost half a far detector module --, and further improvements are present for the atmospheric angle, $\theta_{23}$, and the accuracy of the mass splitting, $\Delta m_{31}^2$.

We have also studied the sensitivity to the neutrino mass ordering, but have found that it benefits less from the improved energy resolution because even without such improvement, DUNE will be able to determine the mass ordering at $5\sigma$ with just \SI{100}{kt\,MW\,years} of exposure \cite{dunecollaboration2021low}.

\textbf{Dependence on neutrino--nucleus cross section modeling.---}An important caveat to the results presented above is that the DNN has been trained on Monte Carlo events, and it is known that current modeling of neutrino--nucleus interactions exhibits considerable discrepancies when compared to experimental data~\cite{Dolan:2019bxf, electronsforneutrinos:2020tbf, MINERvA:2020zzv, CLAS:2021neh, Brdar:2021ysi}.
Although short-baseline detectors can be used to tune event generators, these tunes are somewhat \emph{ad hoc} and what works for one experimental analysis may not be appropriate for another~\cite{Coyle:2022bwa}.

\begin{figure}
    \centering
    \includegraphics[width=\columnwidth]{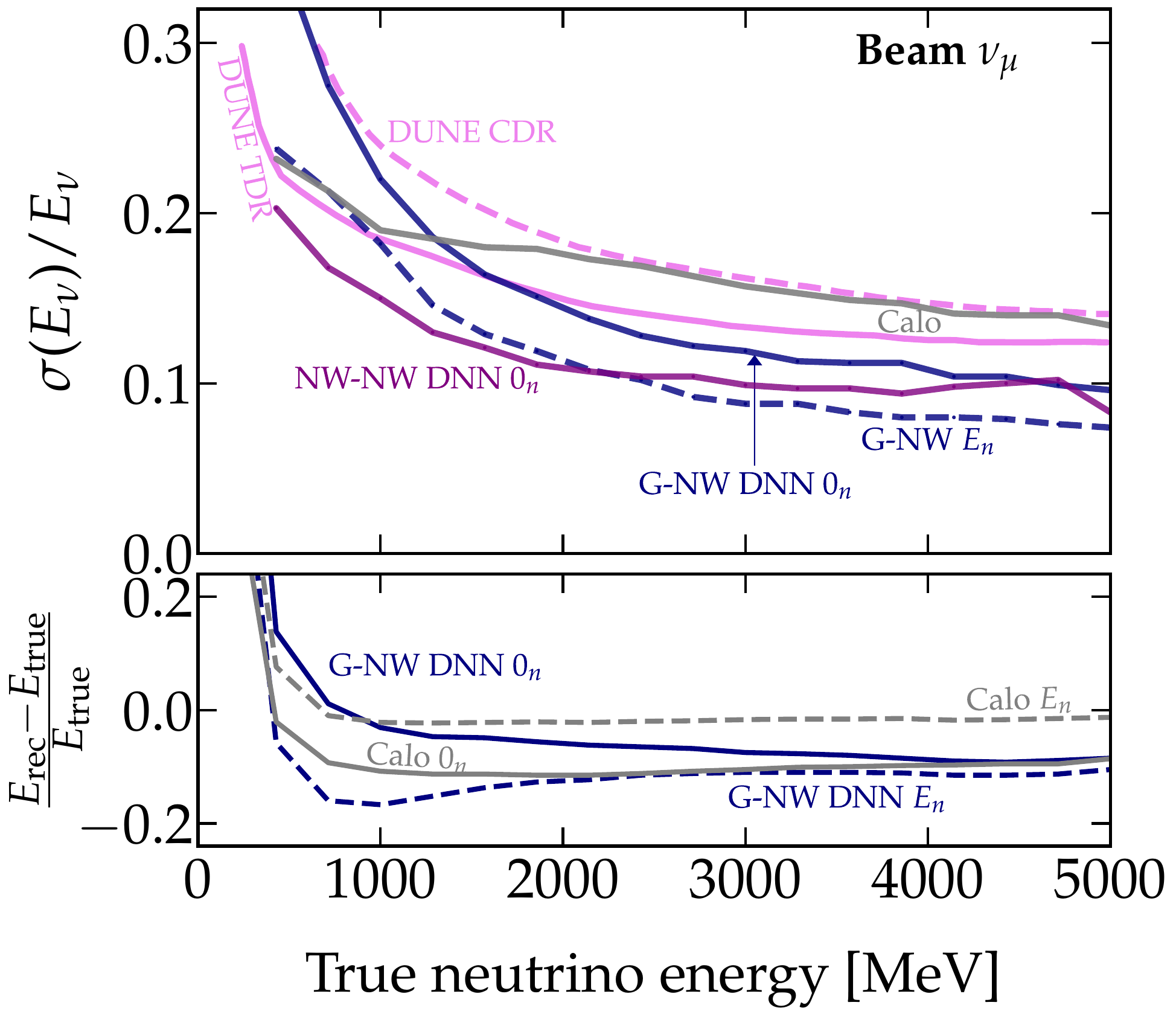}
    \caption{\emph{Top}: fractional neutrino energy resolution, $\sigma(E_\nu)/E_\nu$ from a DNN trained on NuWro events, but applied to GENIE events (blue).
    We compare to the resolutions quoted in the DUNE CDR and TDR (cyan), and to the performance of the calorimetric method.
    \emph{Bottom}: Energy reconstruction bias for the DNN trained on the ``wrong'' neutrino--nucleus interaction model, and for the calorimetric method.
    \emph{Mismodeling of neutrino--nucleus interactions can be a significant limitation to neural network-based and calorimetric reconstruction algorithms alike, leading to biased results.}}
    \label{fig:dnn-robust}
\end{figure}

To obtain a qualitative understanding of the impact that uncertainties associated with neutrino--nucleus interactions have on the performance of our DNN, we have applied our network trained on NuWro~$21.09$ events to mock data generated with GENIE~$3.4.0$. 
In the upper panel of \cref{fig:dnn-robust}, which should be compared to \cref{fig:beam}, we present the DNN's resulting energy resolution (dark blue) and compare it our previous results.
We see that the network underperforms at low energies, while retaining much of its power above about \SI{1.5}{GeV}.
Most likely this is due to differences on how these generators approach intranuclear cascades and low energy outgoing nucleons.

More importantly, the lower panel of \cref{fig:dnn-robust} reveals a \emph{bias} in the neutrino energies reconstructed by the DNN trained with the ``wrong'' neutrino--nucleus interaction model.
The bias is of order 10\% across most of the energy spectrum and, interestingly, is worse when the network is aware of neutron kinematics.
Importantly, though, the calorimetric method exhibits a similar bias when neutron kinematics are not available.
In this case, in fact, the DNN still outperforms calorimetry by a large margin up to energies of several GeV.
Only when neutrons can be reconstructed, the calorimetric method appears robust.
We conclude that mismodeling of neutrino--nucleus interactions remains a potential problem for both traditional and DNN-based neutrino energy reconstruction methods.
This reaffirms the urgency of improving our understanding of the physics of neutrino--nucleus interactions.

\vspace{0.5cm}

 \section*{Acknowledgments}
We thank Ryan Nichol and Linyan Wan for useful discussions.
PANM is supported by Fermi Research Alliance, LLC under Contract No. DE-AC02-07CH11359 with the U.S. Department of Energy, Office of Science, Office of High Energy Physics.
JK would like to acknowledge support through a Fermilab Neutrino Physics Center Fellowship during part of this work. IMS is supported by STFC grant ST/T001011/1.

\clearpage
\bibliographystyle{JHEP}
\bibliography{main}

\end{document}